\documentclass{appolb}
\usepackage{graphicx}
\def\e{\epsilon}
\def\mat#1{\bm{#1}}

\def\k{\vec{k}}

\def\komma{\; ,\;}
\def\punkt{\; .\;}

\def\expect#1{\langle #1 \rangle}

\def\ket#1{|#1\rangle}

\def\mat#1{\underline{\underline{#1}}}
\def\intinf{\int_{-\infty}^{\infty}}
\def\iwn{i\omega_n}

\begin{document}
\title{
Composite spin and orbital triplet superconductivity
\thanks{Presented at the Strongly Correlated Electron Systems 
Conference, Krak\'ow 2002}%
}


\author{Frithjof B. Anders
\address{Institut f\"ur Festk\"orperphysik,
Darmstadt University of Technology,  Hochschulstr.~8, 64289 Darmstadt, Germany}
}
\maketitle


\begin{abstract}
We show that the  two-channel Anderson lattice model leads to the
development of an
unconventional superconductivity out of a metallic non Fermi-liquid phase.
It is characterized  by a composite order parameter
comprising of a local spin or orbital degree of freedom bound to
triplet Cooper pairs with an isotropic and a nearest neighbour form
factor. 
The gap function is non
analytic and odd in frequency, and a pseudo-gap develops in the
conduction electron density of states which vanishes as $|\omega|$
close to $\omega=0$. 
\end{abstract}


  
\section{Introduction}
Heavy Fermion \cite{Grewe91} (HF) superconductivity  (SC) has drawn
much attention since the discovery of
superconductivity in CeCu$_2$Si$_2$ \cite{Steglich79} which is likely
characterized by an anisotropic order parameter with  symmetry yet  to be
determined. It became apparent over the last decade that almost all
HF materials are unstable  with respect to magnetic or superconducting
phase transitions, which either compete with each other or can even
coexist as found in uranium based materials. Moreover, superconductivity  in
UBe$_{13}$ develops out of an incoherent metallic phase with a high
resistance at $T_c$ of $\approx 100\mu \Omega cm$. The two-channel Anderson
lattice model,
%
 \begin{eqnarray}
\label{eq:tca-97}
  \hat H  &= &
\sum_{\k\alpha\sigma} \e_{\k}
c^\dagger_{\k\alpha\sigma}c_{\k\alpha\sigma}
+\sum_{i\sigma} E_\sigma X_{\sigma,\sigma}^{(i)}
+\sum_{i\alpha} E_\alpha X_{\alpha,\alpha}^{(i)}
\nonumber\\  &&
 +
\sum_{i\sigma\alpha=\pm 1} \alpha V\left\{
c^\dagger_{i\alpha\sigma}  X_{-\alpha,\sigma}^{(i)}
+ h.c
 \right\}
\komma
\end{eqnarray}
is believed to serve as a possible descriptions of the electronic
properties of UBe$_{13}$  or Pr$^{3+}$ compounds with cubic
symmetry. $\e_{\k}$ is the band dispersion of the two
conduction bands $\alpha$ coupling to the localized doublets via hybridization
$V$. $X^{(i)}_{\alpha,\beta}$ are the usual Hubbard operators describing
local $f$-states at each lattice side $i$. One doublet is assoziated with
a quadrupolar charge distribution $(X_{\alpha,\alpha})$ and one
carried a spin $(X_{\sigma,\sigma})$. 
Its paramagnetic  phase is characterized by a large 
residual resistivity and entropy, and ill defined electronic
quasi-particles. 
Fermi liquid  physics is restored by cooperative
ordering or applied magnetic field \cite{AndersJarCox97}.

\section{Composite order parameter}

It was observed that the local two-particle-particle irreducible vertex
has a $1/\omega$ singularity and is odd with respect to the incoming
and outgoing frequencies \cite{Details}. This is related to the fact
that correlations are induced through consecutive hybridization
processes on the $f$-shell due to the strong on-site Coulomb
repulsion. Therefore, odd-frequency superconductivity \cite{Heid95} is
favoured in this model. Assuming an order parameter with $\Gamma_1$
symmetry, the order parameter must have spin singlet, channel single
(SsCs) symmetry or spin triplet, channel triplet (StCt) symmetry. Only
for the latter case, a stable solution was found. The order parameter
in the StCt sector  
\begin{eqnarray}
 O_{ij} &= &
\mbox{sign}( E_\sigma-E_\alpha)\left[ \expect{s_i P^{s,t}_{j}}
- \expect{\tau_j P^{t,s}_{i}}\right]
\komma
\label{eqn:stct-order-param}
\end{eqnarray}
correlates a local $f$-spin (quadrupolar moment)  $s_i$ ($\tau_i$) -
$i,j =x,y,z$ - component with a itinerate Cooper pair 
$\vec{P}^{s,t}$ ($\vec{P}^{t,s}$) in the  SsCt (StCs) sector:
\begin{eqnarray}
   \vec{P}^{s,t} & =&
\frac{1}{N} \sum_{\k} S(\k) \vec{\psi}^T(\k)i\mat{\sigma}_y
 i\mat{\tau}_{y}\mat{\vec{\tau}}  \vec{\psi}(-\k)
\\
\vec{P}^{t,s}  
& = & 
\frac{1}{N} \sum_{\k} S(\k)
  \vec{\psi}^T(\k)i\mat{\tau}_y
 i\mat{\sigma}_{y}\mat{\vec{\sigma}}\; \vec{\psi}(-\k)
\punkt
\end{eqnarray}
The form factor 
$S(\k)$ transforms as $\Gamma_1$, and $\vec{\psi}(\k)$ is a
bi-spinor in spin and channel space. $O_{i,j}$
is invariant  with respect to exchanging 
the orbital and the magnetic doublet $\ket{\alpha}$ and $\ket{\sigma}$
and a simultaneous particle-hole transformation of the conduction
electrons. Hence, the occurrence of superconductivity is independent of
the character of the local ground state of the local ion.
We investigate the superconducting phase within
the dynamical mean field theory (DMFT) by self-consistently
calculating the anomalous self-energy under the presents of a
Cooper-pair field.

\section{DMFT in the superconduction phase}

The Nambu Green function is an $8\times 8$ matrix in  spin and
orbital space. Assuming no directional coupling between spin and
orbital degrees of freedom, the anomalous  $4\times 4$ self-energy
matrix may be written as
$g(z,\k) \mat{\sigma}_2\mat{\tau}_2
\left[\vec{n_s}\vec{\mat{\sigma}}\vec{n_c}\vec{\mat{\tau}} \right]$
  where $\vec{n_s}$ and
$\vec{n_c}$ are constant  unity vectors in spin and channel  space,
and the  amplitude function  $g(z,\k) $
has to be odd in frequency. This  reduces the full problem to a standard
$2\times 2$  size for $g(z,\k)$ and the diagonal self-energy $\Sigma$.
In order to derive  DMFT equations with a purely local
self-energy matrix $\mat{\Sigma}_c(z)$, the anomalous
self-energy is restricted to isotropic pairs, e.\ g.\ $g(z,\k) =g(z)$.
$\mat{\tilde G}_c(z)$ denotes the  medium matrix in which the
effective impurity is embedded. It is related to a generalized
Anderson width matrix through
$  \mat{\Delta}(z) = V^2\mat{\sigma}_2 \,\mat{\tilde
    G}_c(z) \,\mat{\sigma}_2
$
and has normal and anomalous components describing quasi-particle
propagation and pair-creation and annihilation, respectively. The
band electron self-energy $
\mat{\Sigma}_c(z) = \mat{T} \left[\mat{1} + \mat{\tilde G} \mat{T} \right]^{-1}
$
is determined by the local $T$-matrix $ \mat{T}$ whose
diagonal elements  are given by the local
quasi-particle scattering matrices $\mat{T}_{11}(z)= V^2 G_f(z)$,
and its anomalous contribution stems from Cooper pairs scattered on
the local shells.


As an impurity solver for effective site problem of the DMFT,
the NCA \cite{Grewe83} was extended to the
superconducting phase. The additional terms $\Delta \Sigma_\gamma(z)$
to the self-consistency equation for the local ionic propagators 
are generated by the anomalous media in the SC phase.
\begin{eqnarray}
\Delta  \Sigma_\alpha(z) & =& 
\sum_\sigma |V|^4\intinf\intinf d\e d\e' \rho_f(\e)\rho_f(\e)f(-\e')
\\
&& \cdot P_\sigma(z+\e) P_\alpha(z+\e-\e') P_\sigma(z-\e')
\nonumber
\\
\Delta  \Sigma_\sigma(z) & =& 
\sum_\alpha |V|^4\intinf\intinf d\e d\e' \rho_f(\e)\rho_f(\e)f(-\e')
\\
&& \cdot P_\alpha(z-\e) P_\sigma(z+\e'-\e) P_\sigma(z+\e')
\nonumber
\end{eqnarray}
where $\rho_f(\e) = \Im m [ \mat{\tilde G}]_{12}(\e-i\delta)/\pi$ is
the spectral function of the anomalous media. The anomalous local
$T$-matrix is also obtained by the same additional diagram of the
generating functional and reads
\begin{eqnarray}
  T_{12}(\iwn) &=& \intinf d\e \rho_f(\e)f(-\e) \oint_{{\cal C}} \frac{e^{-\beta
      z}}{2\pi i Z_f} P_\alpha(z) \\
&& \cdot P_{\sigma}(z+\iwn) P_\alpha(z+\iwn
    -\e) P_\sigma(z-\e)
\nonumber
\komma
\end{eqnarray}
where $Z_f$ is the partition function of the effective site, and the
contour {\cal C} encircles all singularites of the integral kernel. These
coupled double integral equations are numerically solved using  similar 
 techniques as  developed for the Post-NCA equations \cite{Anders95}.

\begin{figure}[!ht]
\begin{center}
\includegraphics[width=0.6\textwidth]{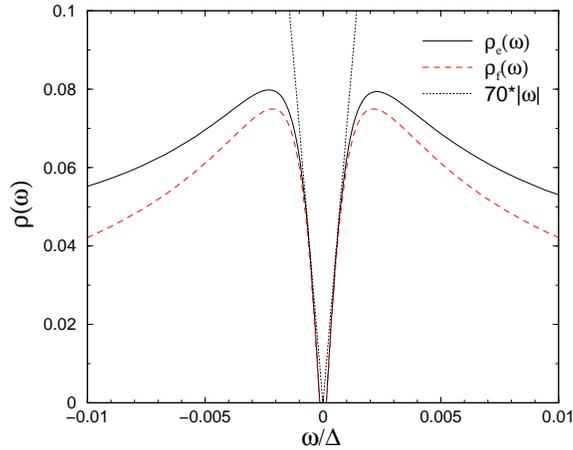}
\end{center}
 \caption{Spectral functions of the quasi-particle and the
      anomalous Green function  in DMFT(NCA) in the superconducting
      phase for $1-T/T_c\ll 1$ in the vicinity of $\omega-\mu=0$.    
      The dotted curve shows a  fit with $a*|\omega|$. 
      Parameters:  $\rho(\e)= exp(-[(\e-\mu)/W]^2)/(\sqrt{\pi}W)$
      conduction band DOS, band width $W=10\Delta$, $E_\sigma
      -E_\alpha = -2\Delta$, $\Delta = V^2\pi \rho(\mu)$,
}
\label{fig-1}
\end{figure}

We calculated the self-consistent DMFT solution of the spectral
functions close to $T_c$ which are displayed in
Fig.~\ref{fig-1}. The linear frequency dependency of the quasi-particle
and anomalous Green function very close to
$\omega=0$, similar to $d$-wave superconductivity, is clearly
visible. The opening of a pseudo-gap indicates 
a decrease of the energy in the superconducting phase, and, hence, its
thermodynamical stability. Details of the filling dependence and the
phase diagram are 
beyond the scope of this paper and will be
published elsewhere \cite{Details}.

\end{document}